\documentclass[sigplan]{acmart}

\usepackage{tabularx}

\usepackage{graphicx}
\usepackage{subcaption}

\newcommand{\hstore}{{\small \sf H-Store}}

\newcommand{\lads}{{\small \sf LADS}}

\newcommand{\calvin}{{\small \sf Calvin}}

\newcommand{\bohm}{{\small \sf BOHM}}

\newcommand{\tpart}{{\small \sf T-Part}}

\AtBeginDocument{%
  \providecommand\BibTeX{{%
    \normalfont B\kern-0.5em{\scshape i\kern-0.25em b}\kern-0.8em\TeX}}}

\begin{document}

\title{A Queue-oriented Transaction Processing Paradigm}
\author{Thamir M. Qadah}
\orcid{0000-0003-0754-0504}
\authornote{The author is co-advised by Prof. Mohammad Sadoghi}
\affiliation{Exploratory Systems Lab}
\affiliation{%
  \institution{School of Electrical and Computer Engineering, Purdue University, West Lafayette}
}
\additionalaffiliation{%
	\institution{Umm Al-Qura University, Makkah, Saudi Arabia}
}
\email{tqadah@purdue.edu}

\begin{abstract}
Transaction processing has been an active area of research for several decades. A fundamental characteristic of classical transaction processing protocols is non-determinism, which causes them to suffer from performance issues on modern computing environments such as main-memory databases using many-core, and multi-socket CPUs and distributed environments. Recent proposals of deterministic transaction processing techniques have shown great potential in addressing these performance issues. In this position paper, I argue for a queue-oriented transaction processing paradigm that leads to better design and implementation of deterministic transaction processing protocols. I support my approach with extensive experimental evaluations and demonstrate significant performance gains.
\end{abstract}



\maketitle

\section{Introduction}
Transaction processing is an old-aged problem that has been an active area of research for the past $40$ years\cite{gray_transaction_1992}. Classical transaction processing is characterized as non-deterministic because the final database state cannot be entirely determined by the input database state and the input set of transactions. The output database state acceptable as long as the resulted history of concurrent transaction execution is equivalent to some serial history of execution according to serializability theory.

The goal of transaction processing protocols is to ensure ACID properties and increase the concurrency of executed transactions. Serializable isolation ensures anomaly-free execution. Using other isolation levels (e.g., Read-committed) improves concurrency but is prone to producing anomalies that defy users' intentions and leave the database in an undesirable inconsistent state. 

Due to the non-deterministic nature of classical transaction processing protocols, they suffer from performance issues on modern computing environments such as main-memory databases that use many-core and multi-socket CPUs, and cloud-based distributed environment. In this Ph.D. dissertation, I look into ways to impose determinism to improve the performance of transaction processing in modern computing environments. 

\section{Transaction Processing in Modern Computing Environments}
\label{sec:challenges}
This section describes two major performance issues when running database transactions using non-deterministic transaction processing protocols. In this section, our discussion assumes the requirement of a serializable isolation model.

\subsection{High-contention Workloads}
\label{sec:high-cont}
Under high-contention workloads, non-deterministic transaction processing protocols suffer from high abort rates because their concurrency control algorithms need to ensure serializable histories. Pessimistic concurrency control algorithms, abort transactions to avoid deadlocks, and optimistic concurrency control algorithms abort transactions during the validation phase. Ensuring deadlock-free execution and validating transactions require extensive coordination among concurrent threads executing transactions while guaranteeing serializability. The main research question for this problem is: {\em Is it possible to process high-contended workloads in a concurrency control-free and minimal coordination while ensuring serializability? What is the right abstraction and principles to achieve that?} 

\subsection{Distributed Commit Protocols}
\label{sec:dist-commit}
In distributed transaction processing, agreement protocols introduce significant overhead to the processing because all participant nodes need to agree on the fate of an executed distributed transaction. Achieving this agreement involves multiple rounds of communication messages to be exchanged among participating nodes. 

The state-of-the-art to solve the agreement problem on the fate of transactions in database systems is the two-phase commit protocol (2PC) \cite{gray_notes_1978-1}. In general cases, 2PC is required to ensure atomicity for processing distributed transactions. Note that 2PC by itself does not ensure serializable histories. A distributed concurrency control augments it to guarantee serializable execution of transactions. Therefore, the research questions for this problem are as follows: {\em Can we reduce the cost of commitment in distributed transaction processing protocols? What conditions are needed to avoid using the costly 2PC-based protocol?}

Fortunately, in many useful and practical cases, we {\em can} do away with 2PC. The work on deterministic transaction processing protocols has demonstrated that. The next section describes how determinism is a step toward overcoming this obstacle. However, proposed deterministic transaction processing protocols suffer from in-efficiencies. Another step toward eliminating these in-efficiencies is the proposed queue-oriented paradigm, which addresses the following additional research questions: {\em What is the best way to abstract deterministic transaction processing? Is it possible to provide a unified framework that unifies transaction processing for centralized and distributed transaction processing?} 

\subsection{Potentials and Limitations of Determinism}
\label{sec:det}
Work on deterministic transaction processing protocols has demonstrated great potential for improving the performance of transaction processing systems \cite{abadi_overview_2018}. In distributed transaction processing systems, recently proposed deterministic approaches {\em almost} eliminates the need to perform a costly 2PC protocol \cite{thomson_calvin_2012}. In other words, they rely on commit protocols that minimize overhead for committing a distributed transaction because they perform agreement ahead of time, which avoids aborting transactions for non-deterministic reasons (e.g., deadlocks, validation, or node failures). 

In deterministic databases, the output database state is entirely determined by the input database state and the input set of transactions. Thus, the full knowledge of the read/write set is required to process transactions deterministically, which the main weakness of using deterministic transaction processing protocols. Despite the existence of such limitation, there are commercial offerings that adopt this deterministic philosophy \cite{voltdb_voltdb_2019, faunadb_faunadb_2019}, which indicates that the approach has found practical use cases in practice.

\begin{table*}[t]
	\begin{center}
		\begin{tabularx}{\textwidth}{p{5cm}p{4cm}p{8cm}}
			\hline 
			\textbf{Name} & \textbf{Fragment relation} & \textbf{Notes} \\ 
			\hline 
			
			Data dependency & Same transaction &  dependent fragment required values read by dependee fragment\\ 
			Conflict dependency & Different transactions & framents access the same record  \\ 
			Commit dependency & Same transaction & dependee fragment may abort and dependent fragment updates the database \\ 
			Speculation dependency & Different transactions &  dependent fragment uses data values updated by is an abortable fragment\\ 
			\hline 
		\end{tabularx}
	\end{center}
	\caption{Summary of dependencies in the transaction fragmentation model}
	\label{tab:txn-frag}
\end{table*}

\section{Approach}

Our goal is to process transactions efficiently in modern computing environments with minimal coordination among the threads running in our system. The proposed approach addresses the research questions presented in the previous section. The answer to these questions relies on three principles: transaction fragmentation, deterministic two-phase processing, and priority-based queue-oriented representation of transnational workload. The essence of the approach is to minimize the overhead of transactional concurrency control and coordination across the whole system. A second goal is to provide a unified extensible abstraction for deterministic transaction processing that seamlessly admits various configurations (e.g., speculative execution, conservative execution, serializable isolation, and read-committed isolation). To lay the foundations for describing the queue-oriented transaction processing paradigm, we start by describing the transaction fragmentation model.

\subsection{Transaction Fragmentation Model}
\label{sec:txn-model}
Now, I briefly describe the transaction fragmentation model. For more formal specification of this model, I refer the readers to \cite{qadah_quecc_2018-1}. In this model, a transaction is broken into fragments containing the relevant transaction logic and aborting conditions. A fragment can perform multiple operations on the same record, such as read, modify, and write operations. A fragment can cause the transaction to abort, and in this case, we refer to such fragments as abortable fragments. In  Table \ref{tab:txn-frag}, a summary is provided on the kinds of dependencies that may exist among fragments.

\subsection{Queue-oriented Transaction Processing}

\begin{figure*}[t]
	\centering
	\includegraphics[width=0.9\textwidth]{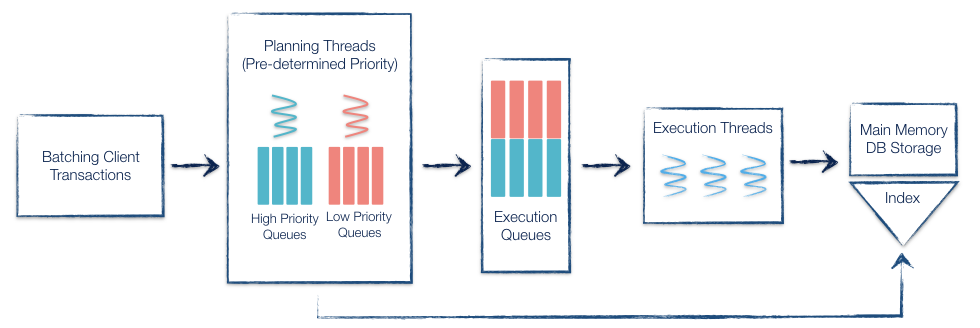}
	\caption{Queue-oriented Transaction Processing Architecture}
	\label{fig:arch}
\end{figure*}

The essence of this paradigm is to process batches of transactions in two deterministic phases. Figure \ref{fig:arch} depicts the basic flow. The first phase is called a planning phase, where threads deterministically create queues tagged with deterministic priorities containing transaction fragments. Dependencies among fragments are not shown in Figure \ref{fig:arch}. The dependency information is maintained in a shared lock-free and thread-safe distributed data structure. In the second execution phase, execution threads receive their assigned queues (filled with fragments) and use the tagged priorities to determine the processing order of queues from different planning threads. At this point, execution threads are not ware of the actual transactions. They are simply executing the logic associated with the fragments in the queues, and obey the FIFO property of queues when processing fragments with conflict dependencies. Processing all queues is equivalent to processing the whole batch of planned transactions and committing them. Other than the necessary communication to resolve dependencies among fragments, no other coordination is needed. 

\textbf{Queue Execution Mechanisms}. The proposed paradigm supports multiple execution mechanisms, such as speculative or conservative. When using speculative execution, additional speculation dependencies occur. Resolving them may cause cascading aborts. Conservative execution, on the other hand, ensures that uncommitted updated are not processed until all abortable fragments complete with aborting, which require additional synchronization and coordination among threads.


\textbf{Isolation Levels}. The queue-oriented paradigm admits read-committed isolation level in addition to serializable isolation. Supporting read-committed isolation with speculative is interesting as it requires maintaining a speculative version and a committed version of records. Other than the storage requirements, the planning phase would create additional queues for read operations. In the execution phase, multiple threads can execute these read operations using committed data. 


\section{Evaluation}
For evaluation, I implemented the queue-oriented processing protocol in ExpoDB \cite{gupta_easycommit_2018, gupta_blockchain_2018-1}. I also ported the state-of-the-art non-deterministic and deterministic protocols into ExpoDB. Using a single test-bed implementation allows apple-to-apple comparison among different protocols. I used industry-standard macro-benchmarks such as YCSB \cite{cooper_benchmarking_2010}, and TPC-C \cite{tpc_tpc-c_2010}. Table \ref{tab:exp} summarizes the experimental results obtained from the centralized implementation running on multi-core hardware with speculative execution. More details is available in \cite{qadah_quecc_2018-1} for our centralized implementation. Furthermore, Table \ref{tab:exp} reports results for our distributed implementation against state-of-the-art distributed deterministic transaction processing protocol. The key performance metrics for evaluating transaction processing protocols are throughput and latency. 

Another criterion for evaluating this paradigm is the applicability and broader impact. For this criterion, the queue-oriented paradigm scores high because it is the first deterministic transaction processing paradigm that allows different execution models and isolation levels. It also has the potential to guide implementations that improves blockchain systems.  

\begin{table*}
	
	\begin{tabularx}{\textwidth}{p{2cm}p{3cm}p{2cm}p{2.5cm}p{3cm}}
		\hline 
		\textbf{Environment} &  \textbf{Compared Protocols} & \textbf{Throughput improvement} & \textbf{Macro-benchmark} & \textbf{Notes} \\ 
		\hline 
		Centralized & (deterministic) H-Store \cite{kallman_h-store_2008}  & two-orders of magnitude & YCSB & Multi-partition workload \\ 
		\hline 
		Distributed  & (deterministic) Calvin \cite{thomson_calvin_2012} & $22\times$ & YCSB & Low-contention workload (Uniform access) \\ 
		\hline 
		Centralized & (non-deterministic) Cicada \cite{lim_cicada_2017}, TicToc \cite{yu_tictoc_2016}, Foudus \cite{kimura_foedus_2015}, Ermia \cite{kim_ermia_2016}, Silo \cite{tu_speedy_2013}, 2PL-NoWait \cite{yu_staring_2014} & $3\times$ & TPC-C & High-contention workload (1 warehouse) \\ 
		\hline 
	\end{tabularx} 
	\caption{Experimental results using TPC-C and YCSB for the centralized implementation of queue-oriented paradigm \cite{qadah_quecc_2018-1}, and a distributed deterministic database.}
	\label{tab:exp}
\end{table*}

\section{Related Work}
\label{sec:rw}
The related work to my Ph.D. dissertation falls into two categories. 
In the first category, many centralized deterministic transaction processing protocols are proposed. \lads\ by Yao et al. \cite{yao_exploiting_2016} creates multiple sub-graphs representing transaction dependencies of a batch of transactions, and execute these transactions according to the dependency sub-graphs. The main issue with this approach is the graph-based processing is not efficient. Using a different graph-based approach, Faleiro et al. \cite{faleiro_high_2017} process transactions deterministically and introduces the notion of ``early write visibility'' which allows transactions to read uncommitted data safely. In our approach, we use queues of transaction fragments with different dependency semantics that allows us to process transactions more efficiently when compared to a purely graph-based approach. \bohm\ \cite{faleiro_rethinking_2015} started re-thinking multi-version concurrency control for deterministic multi-core in-memory data stores. \bohm{} relies on pessimistic transactional concurrency control while our proposed paradigm avoids transactional concurrency control during execution. Some ideas presented in \cite{faleiro_rethinking_2015, faleiro_high_2017} are complementary to our approach. For example,  our current implementation is single-version but can be extended to multi-version in the future.

In the second category, one of the first proposed distributed deterministic database systems is \hstore\ \cite{kallman_h-store_2008}, which focuses on partitioned workloads. The design of \hstore\ does not lend itself to work well for multi-partition transactional workload because of the partition-level locking mechanism and 2PC. To improve the performance of multi-partition workloads, Jones et al. \cite{jones_low_2010} introduced the idea of speculative execution in \hstore\ while still relying on 2PC as a distributed commit protocol. In contrast to these proposals, the use of speculative execution in the proposed paradigm is different because speculative execution is at the level of fragments. Furthermore, the proposed paradigm does not require 2PC to commit distributed multi-partition transactions.

As mentioned previously, \calvin\ \cite{thomson_calvin_2012} greatly reduces the overhead of distributed transactions because it does not rely on 2PC. Wu et al. proposes \tpart\ \cite{wu_t-part_2016}, which uses the same fundamental design as \calvin. \tpart\ optimizes the handling remote reads by using a forward-pushing technique at the cost of more complex scheduling that involves solving a graph-partitioning problem. The key characteristic of \calvin\ and \tpart\ is that they use thread-to-transaction assignment while our approach uses thread-to-queue assignment. Therefore, these systems cannot exploit intra-transaction parallelism within a single node.

\section{Conclusion}

In this paper, I argued for a queue-oriented transaction processing paradigm, which improves the performance of deterministic databases. On-going work includes using this paradigm to design and implement distributed transaction processing with byzantine-fault-tolerance. 

Future work includes using the proposed paradigm to realize a deterministic version of production-ready NewSQL databases such as TiDB \cite{noauthor_tidb_2019}. Moreover, I believe that this paradigm can also improve the performance of blockchain systems. In particular, the queue-oriented paradigm can lead to a design and implementation that improves the performance of the ordering service in HyperLedger Fabric \cite{androulaki_hyperledger_2018}.

\begin{acks}
	I want to thank my co-advisors Prof. Arif Ghafoor, for his continuous support during my Ph.D. journey, and Prof. Mohammad Sadoghi, for his valuable comments that helped me develop the ideas in my thesis. 	
	The author would also like to thank the anonymous referees and Yahya Javed for their valuable comments and helpful suggestions.
	The work is supported in part by a scholarship from Umm Al-Qura University, Makkah, Saudi Arabia. 
\end{acks}

\bibliographystyle{ACM-Reference-Format}
\bibliography{database_zotero.bib,blockchain.bib}
\end{document}